\documentclass{iaus}
\usepackage{graphicx}

\newcommand{\gray}{$\gamma$-ray}
\newcommand{\grays}{$\gamma$-rays}

\newcommand{\hi}{H {\sc i}}
\newcommand{\hii}{H {\sc ii}}

\newcommand{\hone}{\mbox{$\mathcal{H}_1$}}
\newcommand{\htwo}{\mbox{$\mathcal{H}_2$}}
\newcommand{\psra}{\mbox{PSR~J0540$-$6919}}
\newcommand{\psrb}{\mbox{PSR~J0537$-$6910}}
\newcommand{\ra}{\mbox{$\alpha_{\rm J2000}$}}
\newcommand{\dec}{\mbox{$\delta_{\rm J2000}$}}
\newcommand{\apj}{Astrophys.\ J.}
\newcommand{\apjl}{Astrophys.\ J.\ Lett.}
\newcommand{\apjs}{Astrophys.\ J.\ Supp.}
\newcommand{\aap}{Astron.\ \& Astrophys.}

\newcommand{\arnps}{Ann.\ Rev.\ Nuc.\ Part.\ Sci.\ } 
\newcommand{\pubjournal}[5]{#4, #1, #2, #3}

\newcommand{\fermilatlong}{{\it Fermi} Large Area Telescope}
\newcommand{\fermilat}{{\it Fermi}-LAT}

\title[~~Cosmic-Ray Induced Diffuse Emissions from Local Group Galaxies] %% give here short title %%
{Cosmic-Ray Induced Diffuse Emissions from the Milky Way and Local Group Galaxies}

\author[Troy A. Porter]   %% give here short author list %%
{Troy A. Porter}

\affiliation{Hansen Experimental Physics Laboratory \\
  and \\
  Kavli Institute for Particle Astrophysics and Cosmology \\
  Stanford University, Stanford, USA \\
  email: {\tt tporter@stanford.edu}
}

\pubyear{2011} %% 
\volume{284}  %% insert here IAU Symposium No.
\pagerange{1--12}
\jname{The Spectral Energy Distribution of Galaxies}
\editors{R.J. Tuffs \&  C.C.Popescu, eds.}
\begin{document}

\maketitle

\begin{abstract}
Cosmic rays fill up the entire volume of galaxies, providing an important 
source of heating and ionisation of the interstellar medium, and may play a
significant role in the regulation of star formation and galactic evolution.
Diffuse emissions from radio to high-energy \grays{} ($> 100$ MeV) arising 
from various interactions between cosmic rays and the interstellar medium, 
interstellar radiation field, and magnetic field, are currently the best way
to trace the intensities and spectra of cosmic rays in the Milky Way and 
other galaxies.
In this contribution, I describe our recent work to model the full spectral
energy distribution of galaxies like the Milky Way from radio to \gray{} 
energies.
The application to other galaxies, in particular the Magellanic 
Clouds and M31 that are detected in high-energy \gray{s} by the 
Fermi-LAT, is also discussed.

\end{abstract}

\firstsection % if your document starts with a section,
\section{Introduction}
The luminosity of a star-forming galaxy like the
Milky Way (MW) 
is dominated by the relatively narrow frequency range of the spectral
energy distribution (SED) from the 
ultraviolet (UV) to far infrared (FIR), which is due to 
stellar emission and dust reprocessing in the interstellar medium (ISM).
Related to the birth and death of massive stars, cosmic rays (CRs) are 
pervasive throughout the ISM (see, e.g.,~\cite{Strong2007}~for a recent 
review). 
The diffuse emissions arising from various interactions between CRs and 
the ISM, 
interstellar radiation field (ISRF -- the UV--FIR component of the galactic 
SED), and magnetic field span radio frequencies
to high-energy \grays{} ($>$100 MeV), but at a lower level of intensity 
compared to the stellar and dust component.
These broadband emissions are currently 
the best way to trace CR intensities and spectra throughout 
the MW and other galaxies.
Gamma rays are particularly useful in this respect because this energy 
range gives access to the dominant hadronic component
in CRs via the observation of $\pi^0$-decay radiation produced by 
CR nuclei inelastically colliding with the interstellar gas.
Understanding the global energy budget of processes related to the injection
and propagation of CRs, and how the energy is distributed across the 
electromagnetic spectrum, is essential to interpret the 
radio/far-infrared relation (\cite{Helou1985,Murphy2006}), galactic 
calorimetry (e.g.,~\cite{Volk1989}), and predictions of 
extragalactic backgrounds (e.g.,~\cite{Thompson2007,Murphy2008}), and for
many other studies.

\section{Broadband Spectral Energy Distribution for the Milky Way}
\begin{figure}[htb]
% \vspace*{-2.0 cm}
\begin{center}
 \centerline{\includegraphics[width=2.4in,height=2.2in]{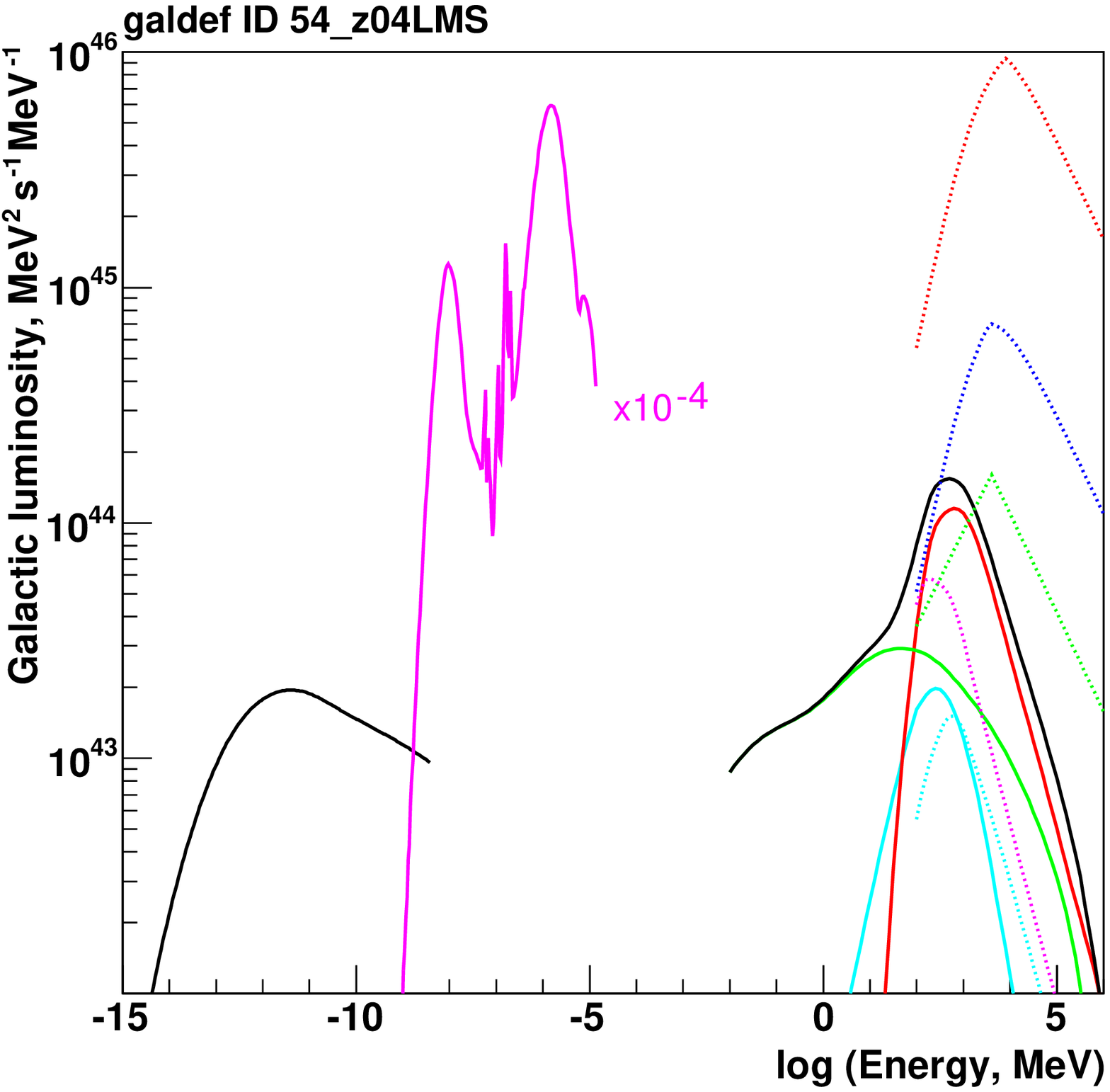}
 \includegraphics[width=2.8in,height=2.2in]{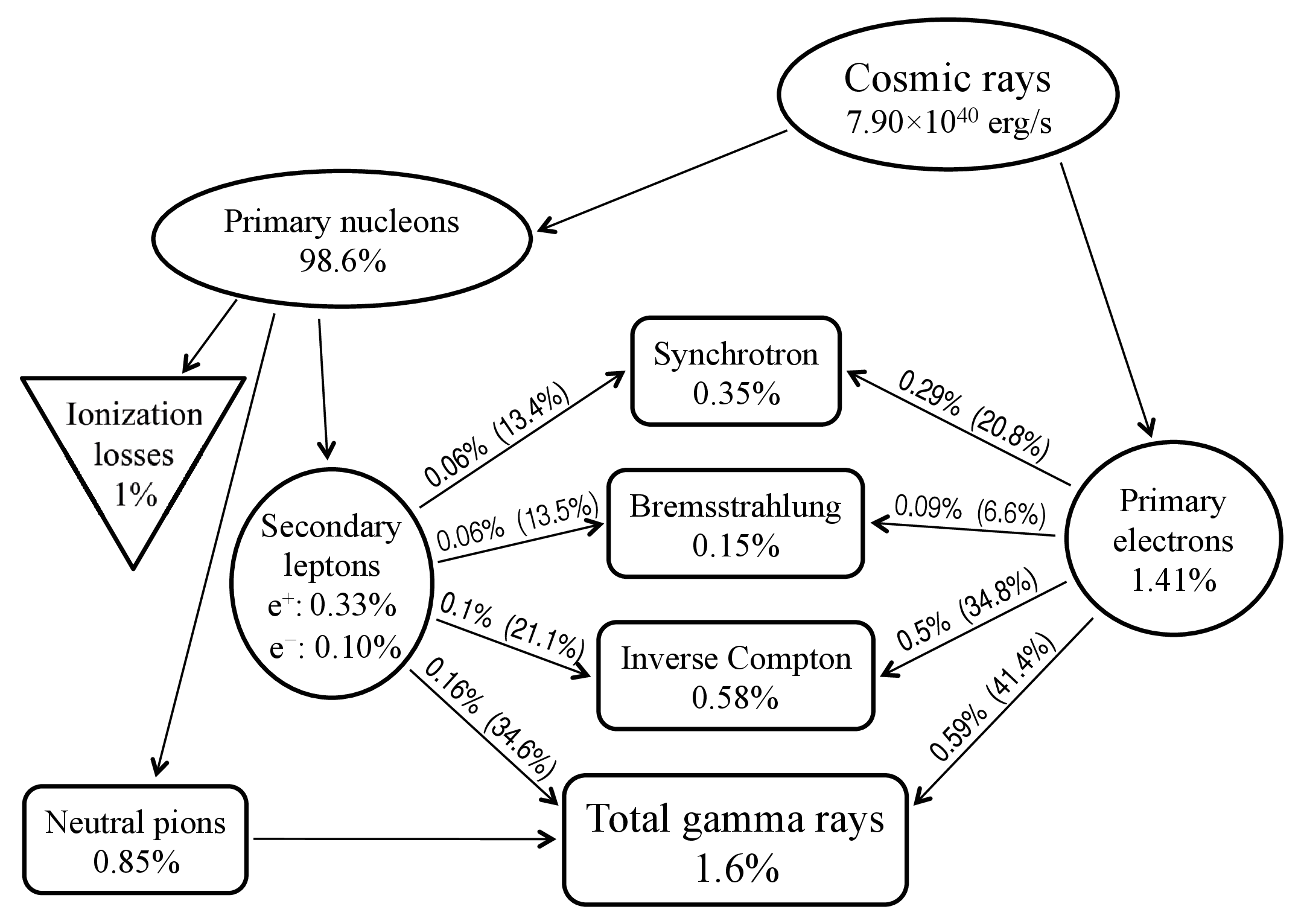}}
% \vspace*{-1.0 cm}
 \caption{{\it Left:} Global CR-induced luminosity spectra of the MW 
for a diffusive-reacceleration CR propagation model with $z _h = 4$ kpc. 
Line styles: ISRF, including optical and 
infrared {\it scaled by factor $10^{-4}$} (magenta solid) and 
components for propagation model --  
Cosmic rays (dotted lines), 
protons (red), helium (blue), primary electrons (green), 
secondary electrons (cyan), secondary positrons (magenta); 
CR-induced diffuse emissions (solid lines), 
IC (green), bremsstrahlung (cyan), $\pi^0$-decay (red), 
synchrotron (black, left side of figure), total (black, right side of figure).
{\it Right:} Luminosity budget of the MW for the CR propagation model
shown in the left panel.
The percentage figures
are shown with respect to the total injected 
luminosity in CRs, $7.9\times10^{40}$ erg s$^{-1}$. 
The percentages in brackets show the values relative to the luminosity of 
their respective lepton populations 
(primary electrons, secondary electrons/positrons).}
   \label{fig1}
\end{center}
\end{figure}

The MW is the best studied non-AGN dominated star-forming 
galaxy, and the only galaxy that direct measurements of CR intensities and 
spectra are available.
However, because of our position inside, the derivation of 
global properties is not straightforward and requires detailed 
models of the spatial distribution of the emission.
In recent work by \cite{Strong2010} using the 
GALPROP (e.g.,~\cite{SMR2000,Moskalenko2002}~; also see 
http://galprop.stanford.edu) and 
FRaNKIE (Fast Radiative Numerical Kode for Interstellar Emission -- 
\cite{Porter2008}) codes, we have calculated the broadband SED of the MW for
the first time.
Emission by stars and dust is included, together with the diffuse 
emissions for different CR propagation models consistent with local CR data. 
Figure~\ref{fig1} (left) shows the broadband luminosity 
spectrum of the Galaxy, including the input luminosity 
for CRs for a 4 kpc halo using a diffusive-reacceleration CR propagation 
model.
\cite{Strong2010} considered a range of halo sizes (2--10 kpc).
For this range, the relative decrease in the injected CR proton and helium 
luminosities is $\sim10$\%.
For smaller halo sizes, the CRs escape quicker requiring more injected power 
to maintain the local CR spectrum.
In addition, for larger halo sizes CR sources located at further distances
can contribute to the local spectrum, which is the normalisation condition,
hence less power is required.
Contrasting with the CR nuclei, the injected 
primary CR electron luminosity {\it increases} 
with $z_h$, which is required 
to counter the increased inverse-Compton (IC) energy losses in 
the halo from the longer escape time.

Figure~\ref{fig1} (right) illustrates the detailed energy budget for the 
propagation model shown in Fig.~\ref{fig1} (left).
The energy channelled into \gray{} and other secondary production 
from the CR nuclei component is only a very small fraction 
of the total power injected in these particles.
For the model shown, the total of synchrotron, IC, and 
bremsstrahlung luminosities for the primary electrons accounts for
over half of the total luminosity injected in these particles.
Note that it is the IC emission at \gray{} energies 
that is responsible for the majority of 
the energy losses, not synchrotron radiation as usually assumed.
Including the contribution by secondary electron/positron production, 
approximately half of the total output in \grays{} is provided by these 
particles, even though they count for only $\sim 2$\% of the injected power.
Non-AGN dominated galaxies like the MW turn out to be very good lepton
calorimeters if {\it all} the energy-loss processes are taken into account.

\section{Gamma Rays from Local Group Galaxies}
Until recently, the MW was the only galaxy that was resolved in high-energy
\grays{}.
However, observations by the \fermilatlong{} (LAT) 
have extended the sample of resolved
star-forming galaxies to include the Magellanic Clouds (\cite{LMCLat,SMCLat}), 
with M31 also being detected (\cite{M31Lat}).
%Measuring the diffuse \gray{} emission of these galaxies is the necessary
%step to compare with SED models for these galaxies.

\begin{figure}[htb]
% \vspace*{-2.0 cm}
\begin{center}
 \centerline{\includegraphics[height=2.2in,width=2.6in]{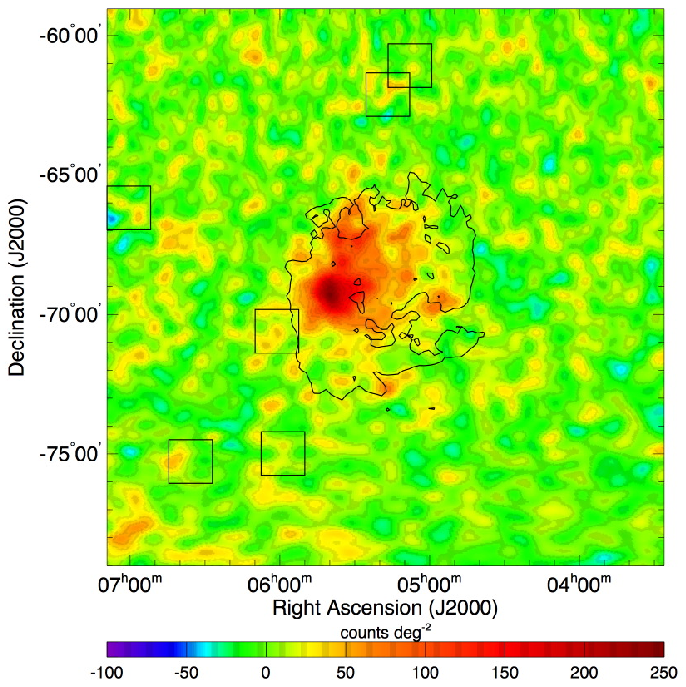}
 \includegraphics[width=2.4in,height=2.2in]{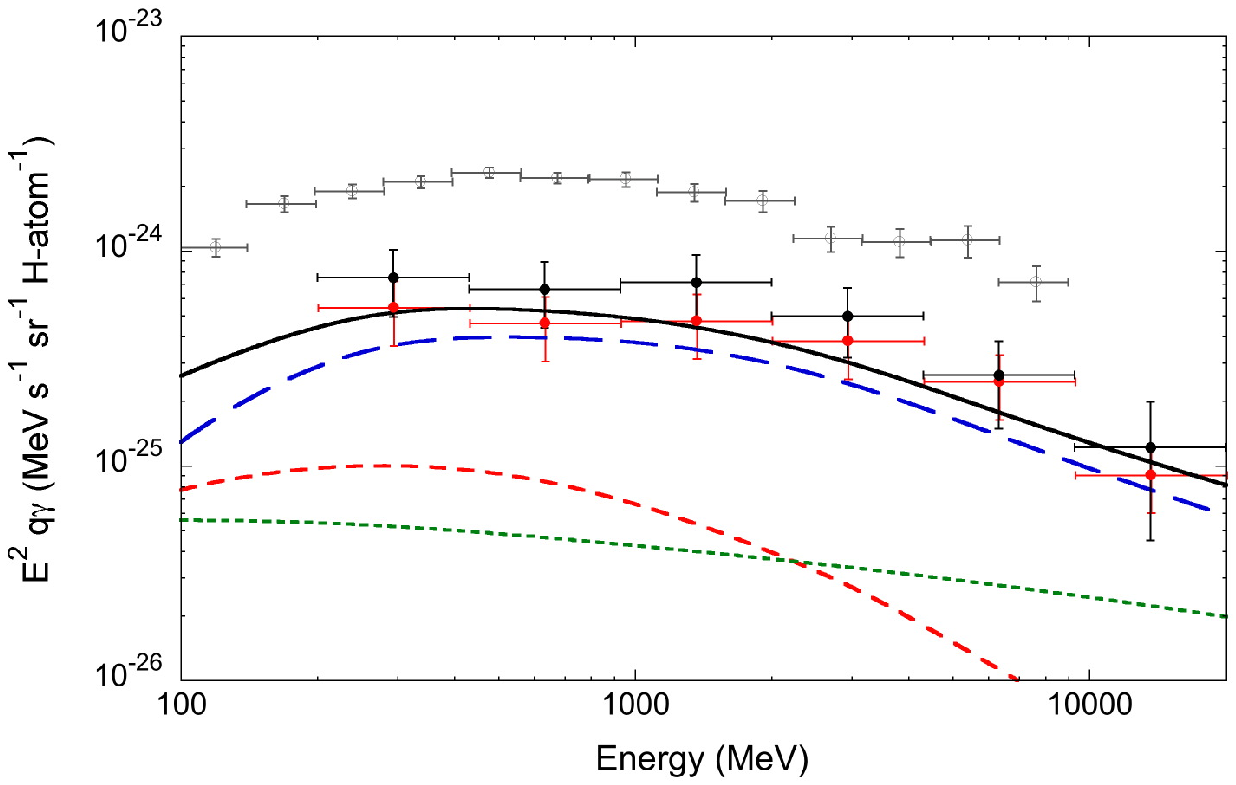}}
% \vspace*{-1.0 cm}
 \caption{{\it Left:}
   Gaussian kernel ($\sigma=0.2^\circ$) smoothed count map for the ROI 
   about the LMC after subtraction of the background model for the energy range 
   200 MeV -- 20 GeV and for a pixel size of $0.1^\circ \times 0.1^\circ$.
   Overlaid is the N(\hi) contour of $1\times10^{21}$ H~cm$^{-2}$ of the LMC to 
   indicate the extent and shape of the galaxy.
   The boxes show the locations of the 6 point sources (background blazars) 
   that were included in the background model.
   {\it Right:} Differential average \gray{} emissivity spectrum for the LMC.
   Data for models \hone\ (black dots) and \htwo\ (red dots) (see text) and
   the local ISM emissivity (\cite{Abdo2009b} -- grey data points) are shown.
   %Error bars for \hone\ and \htwo\ include statistical and 
   %systematic errors and the uncertainties in the total gas mass.
   Model lines: solid, predicted total \gray{} emissivity computed in the 
   framework of a one-zone model for \htwo; long dashed, $\pi^0$-decay;
   short dashed, bremsstrahlung; dotted, IC.
   See \cite{LMCLat} for details.}
   \label{fig2}
\end{center}
\end{figure}

Figure~\ref{fig2} (left) shows the background subtracted counts map for a 
$20^\circ\times20^\circ$ region of interest (ROI) surrounding the LMC.
The remaining feature is extended emission that is spatially confined 
to within the LMC boundaries, which are traced by the iso column density 
contour $N_{\rm H} = 1\times10^{21}$ H~cm$^{-2}$
of neutral hydrogen in the LMC (\cite{kim05}).
The extended \gray{} emission from the LMC can be
resolved into several components.
The brightest emission feature is located near
$(\ra, \dec) \approx (05^{\rm h}40^{\rm m}, -69^\circ15')$,
which is close to the massive star-forming region 30~Doradus (30~Dor)
that houses the two Crab-like pulsars 
\psra\ and \psrb. % \citep{seward84,marschall98}.
Excess \gray{} emission is also seen toward the north and the west of 30~Dor.
These bright regions are embedded within a more extended and diffuse glow that
covers an area of approximately $5^\circ \times 5^\circ$.

%To test specific hypotheses about the spatial distribution of the 
%\gray{} intensity 
%the data were compared 
%to spatial templates that trace the interstellar matter distribution
%in the LMC.
%The reasoning behind this comparison is that \grays{} are expected to primarily
%arise from interactions between CRs and the ISM,  
%and we may thereby determine the CR density variations from the
%\gray{} to gas ratio.

Most of the gas in the LMC is found in the form of neutral hydrogen 
and helium %atomic hydrogen 
%and helium,
%(\cite{staveleysmith03}) 
%while $5\%-10\%$ of the mass is in form of molecular 
%clouds %(\cite{fukui08}), 
with only about $1\%$ of the 
total mass ionised. % (\cite{kennicutt95}).
However, 
the distribution of \hii\ provides the best fit to the observed \gray{}
emission. % among all of the gas components.
It is characterised by a strong emission peak near 30~Dor which is attributed
to the intense ionising radiation of the massive stars in this highly 
active region.
Even accounting for the 30~Dor emission, the \hii\ distribution
is still a significantly better tracer than the neutral gas %components 
of the residual emission.
This contrasts with the MW, where the majority of the diffuse \gray{} 
emission appears to trace the neutral gas.

\cite{LMCLat} tested two hypotheses for the origin of the \gray{} emission
from the LMC: (\hone) all \gray{} emission from the LMC is attributed to 
diffuse emission from CR interactions, and
(\htwo) only emission from a disk-like component arises from CR 
interactions, while the \grays{} from 30~Dor originate from other sources.
The emissivity spectrum derived using these two hypotheses for the LMC
is shown in Fig.~\ref{fig2} (right).
%Using these two hypotheses, the emissivity per hydrogen atom of the LMC
%was obtained, which is shown in Fig.~\ref{fig2} (right) 
%together with the differential emission spectrum calculated
%assuming a one-zone model and the 
%Galactic local emissivity spectrum (\cite{Abdo2009b}).
The integrated emissivity $> 100$ MeV for the LMC is $\sim2-4$ times lower
than the locally derived emissivity (also shown in the figure).
Similar analysis for the SMC (\cite{SMCLat}) and M31 (\cite{M31Lat}) derive
emissivities $> 100$ MeV that are $\sim 6-7$ and $\sim 2$ times 
lower, respectively.
Assuming that the proton-to-electron ratio in these galaxies 
above a few tens of GeV is similar to that in the MW and that all of the
emission is from diffuse processes, these values point to 
a general picture where the intensity of the emission is controlled by
the power injected by CR sources scaling approximately with the relative
star-formation rates in the different systems.
Future work combining the \gray{}, radio, and other data will result 
in broadband SEDs of these galaxies
spanning more than 20 decades in frequency, which can be used to 
investigate a variety of
phenomena related to how energy is injected and cycled within the ISM of
galaxies other than the MW.

%\section{Summary}
%I have described recent work to model the 
%SEDs of star-forming galaxies like
%the MW from radio to \gray{} energies. 
%These calculations include the stellar and dust emission, as well as that from 
%CRs interacting in the ISM.
%With the measurements made of the \gray{} emission of the Magellanic Clouds 
%and M31, it is now possible to construct SEDs that are truly broadband.

\section{Acknowledgements}
The \fermilat\ Collaboration acknowledges support from a number of 
agencies and institutes for both development and the operation of the LAT 
as well as scientific data analysis. These include NASA and DOE in the 
United States, CEA/Irfu and IN2P3/CNRS in France, ASI and INFN in 
Italy, MEXT, KEK, and JAXA in Japan, and the K.~A.~Wallenberg Foundation, 
the Swedish Research Council and the National Space Board in Sweden. 
Additional support from INAF in Italy and CNES in France for science 
analysis during the operations phase is also gratefully acknowledged.
GALPROP development is supported via NASA Grant Nos.~NNX10AE78G and~NNX09AC15G.


\begin{thebibliography}{}

%\bibitem[Amari \etal\ (1995)]{Amari_etal95}
%{Amari, S., Hoppe, P., Zinner, E., \& Lewis R.S.} 1995,
%\textit{Meteoritics}, 30, 490 

%\bibitem[Abdo et al.(2009a)]{Abdo2009a}
%  Abdo, A.~A., et al., \pubjournal{\prl}{102}{181101}{2009}{}

\bibitem[Abdo~et~al.~2009]{Abdo2009b}
  Abdo,~A.~A., Ackermann,~M., Ajello,~M.,~et~al., \pubjournal{\apj}{703}{1249}{2009}{}

%\bibitem[Abdo et al.(2009c)]{Abdo2009c}
%  Abdo, A.~A., et al., \pubjournal{\prl}{103}{251101}{2009}{}

%\bibitem[Abdo et al.(2010a)]{Abdo2010a}
%  Abdo, A.~A., et al., \pubjournal{\apj}{710}{133}{2010}{}

%\bibitem[Abdo et al.(2010b)]{Abdo2010b}
%  Abdo, A.~A., et al., \pubjournal{\prl}{104}{101101}{2010}{}

\bibitem[Abdo~et~al.~2010]{LMCLat}
  Abdo,~A.~A., Ackermann,~M., Ajello,~M.,~et~al., \pubjournal{\aap}{512}{7}{2010}{}

\bibitem[Abdo~et~al.~2010]{M31Lat}
  Abdo,~A.~A., Ackermann,~M., Ajello,~M.,~et~al., \pubjournal{\aap}{523}{L2}{2010}{}

\bibitem[Abdo~et~al.~2010]{SMCLat}
  Abdo,~A.~A., Ackermann,~M., Ajello,~M.,~et~al., \pubjournal{\aap}{523}{46}{2010}{}

%\bibitem[Ando \& Pavlidou(2009)]{Ando2009}
%  Ando, S. \& Pavlidou, V., \pubjournal{\mnras}{400}{2122}{2009}{}

%\bibitem[Binney \& Merrifield(1998)]{Binney1998}
%  Binney, J. \& Merrifield, M., {\it Galactic Astronomy} 
%Princeton, NJ : Princeton University Press, (1998). 
%(Princeton series in astrophysics)

%\bibitem[Cox(2005)]{Cox2005}
%  Cox, D.~P., \pubjournal{\araa}{43}{337}{2005}{}

%\bibitem[de Nolfo et al.(2006)]{deNolfo2006}
%  de Nolfo, G.~A., et al., \pubjournal{\adv}{38}{1558}{2006}{}

%\bibitem[Diehl et al.(2006)]{Diehl2006}
%  Diehl, R., et al., \pubjournal{\nat}{439}{45}{2006}{}

%\bibitem[Dogiel et al.(2002)]{Dogiel2002}
%  Dogiel, V.~A., et al., \pubjournal{\apjl}{572}{157}{2002}{}

%\bibitem[Engelmann et al.(1990)]{Engelmann1990}
%  Engelmann, J.~J., et al., \pubjournal{\aap}{233}{96}{1990}{}

%\bibitem[Ferri\`{e}re(2001)]{Ferriere2001}
%  Ferri\`{e}re, K.~M., \pubjournal{\rmp}{73}{1031}{2001}{}

%\bibitem[Freudenreich(1998)]{Freudenreich1998}
%  Freudenreich, H.~T., \pubjournal{\apj}{492}{495}{1998}{}

%\bibitem[Fukui~et~al.~2008]{fukui08}
%  Fukui,~Y.,~Kawamura,~A.,~Minamidani,~T.
%  \pubjournal{\apjs}{178}{56}{2008}{}

\bibitem[Helou~et~al.~1985]{Helou1985}
  Helou,~G., Soifer,~B.~T., \& Rowan-Robinson,~M., 
  \pubjournal{\apjl}{298}{7}{1985}{}

%\bibitem[Kennicutt~et~al.~1995]{kennicutt95}
%Kennicutt,~R.~C.,~et~al., \pubjournal{\aj}{109}{594}{1995}{} 

%\bibitem[Kent et al.(1991)]{Kent1991}
%  Kent, S.~M., et al., \pubjournal{\apj}{378}{131}{1991}{}

\bibitem[Kim~et~al.~2005]{kim05}
  Kim,~S., Staveley-Smith,~L., Dopita,~M.~A.,~et~al., \pubjournal{\apjs}{143}{487}{2005}{} 

%\bibitem[Lacki et al.(2010)]{Lacki2010}
%  Lacki, B.~C., et al., \pubjournal{\apj}{717}{1}{2010}{}

%\bibitem[{{Lorimer}(2004)}]{Lorimer2004}
%{Lorimer}, D.~R. 2004, in IAU Symposium, Vol. 218, Young Neutron Stars and
%  Their Environments, ed. {F.~Camilo \& B.~M.~Gaensler}, 105

%\bibitem[Moskalenko \& Strong(1998)]{Moskalenko1998}
%  Moskalenko, I.~V. \& Strong, A.~W., \pubjournal{\apj}{493}{694}{1998}{}

\bibitem[Moskalenko~et~al.~2002]{Moskalenko2002}
  Moskalenko,~I.~V., Strong,~A.~W., Ormes,~J.~F.,~et~al., \pubjournal{\apj}{565}{280}{2002}{}

\bibitem[Murphy~et~al.~2006]{Murphy2006}
  Murphy,~E.~J., Helou,~G., Braun,~R.,~et~al., \pubjournal{\apj}{638}{157}{2006}{}

\bibitem[Murphy~et~al.~2008]{Murphy2008}
  Murphy,~E.~J., Helou,~G., Kenney,~J.~D.~P.,~et~al., \pubjournal{\apj}{678}{828}{2008}{}

%\bibitem[Paladini et al.(2007)]{Paladini2007}
%  Paladini, R., et al., \pubjournal{\aap}{465}{839}{2007}{}

\bibitem[Porter~et~al.~2008]{Porter2008}
  Porter,~T.~A., Moskalenko,~I.~V., Strong,~A.~W.,~et~al., \pubjournal{\apj}{682}{400}{2008}{}

%\bibitem[Sironi \& Socrates(2010)]{Sironi2010}
%  Sironi, L. \& Socrates, A., \pubjournal{\apj}{710}{891}{2010}{}

%\bibitem[Socrates et al.(2008)]{Socrates2008}
%  Socrates, A., et al., 
%  \pubjournal{\apj}{687}{202}{2008}{}

%\bibitem[Sodroski et al.(1997)]{Sodroski1997}
%  Sodroski, T.~J., et al., \pubjournal{\apj}{480}{173}{1997}{}

%\bibitem[Staveley-Smith~et~al.~2003]{staveleysmith03}
%Staveley-Smith,~L.,~et~al., \pubjournal{\mnras}{339}{87}{2003}{}

%\bibitem[Strong \& Moskalenko(1998)]{Strong1998}
%  Strong, A.~W. \& Moskalenko, I.~V., \pubjournal{\apj}{509}{212}{1998}{}

\bibitem[Strong~et~al.~2000]{SMR2000}
  Strong,~A.~W.,~Moskalenko,~I.~V.,~\&~Reimer,~O., 
  \pubjournal{\apj}{537}{763}{2000}{}

%\bibitem[Strong et al.(2004a)]{Strong2004a}
%  Strong, A.~W., et al., \pubjournal{\apj}{613}{962}{2004}{}

%\bibitem[Strong et al.(2004b)]{Strong2004b}
%  Strong, A.~W., et al., \pubjournal{\aap}{422}{L47}{2004}{}

\bibitem[Strong~et~al.~2007]{Strong2007}
  Strong,~A.~W., Moskalenko,~I.~V. \& Ptuskin,~V.~S., \pubjournal{\arnps}{57}{285}{2007}{}

\bibitem[Strong~et~al.~(2010)]{Strong2010}
  Strong,~A.~W., Porter,~T.~A., Digel,~S.~W.,~et~al., 
  \pubjournal{\apjl}{722}{58}{2010}{}

\bibitem[Thompson~et~al.~2007]{Thompson2007}
  Thompson,~T.~A.,~Quataert,~E.,~\&~Waxman,~E., 
  \pubjournal{\apj}{654}{219}{2007}{}

\bibitem[V\"{o}lk~1989]{Volk1989}
  V\"{o}lk,~H.~J., \pubjournal{\aap}{218}{67}{1989}{}

%\bibitem[Yanasak et al.(2001)]{Yanasak2001}
%  Yanasak, N.~E., et al., \pubjournal{\apj}{563}{768}{2001}{}

%\bibitem[Yun et al.~(2001)]{Yun2001}
%  Yun, M.~S., et al., \pubjournal{\apj}{554}{803}{2001}{}


\end{thebibliography}
\end{document}